\documentclass[runningheads]{llncs}
\usepackage{appendix}
\usepackage{amsmath}
\usepackage{graphicx}
\usepackage{lineno}
\usepackage{array}
\usepackage{longtable}
\usepackage{todonotes}
\usepackage{graphicx,wrapfig,lipsum}
\usepackage{longtable}
\usepackage{tabularx}
\usepackage{soul}
\usepackage{booktabs}
\usepackage{floatrow}
\usepackage{changes}
\usepackage{hyperref}
\usepackage[square,numbers]{natbib}

\newfloatcommand{capbtabbox}{table}[][\FBwidth]

\begin{document}
%
%
\title{Is Grad-CAM Explainable in Medical Images?}
%
\author{Subhashis Suara\inst{1} \and
Aayush Jha\inst{1} \and
Pratik Sinha\inst{1} \and
Arif Ahmed Sekh$^*$\inst{1}}
\authorrunning{Suara et al.}
%
\institute{XIM University, Bhubaneswar, India\\ 
 \email{\{subhashissuara,aayushjha1990,pratiksinha.cs,skarifahmed@gmail.com\}}
}
\maketitle              
\begin{abstract}
Explainable Deep Learning has gained significant attention in the field of artificial
intelligence (AI), particularly in domains such as medical imaging, where accurate and interpretable machine learning models are crucial for effective diagnosis and treatment
planning. Grad-CAM is a baseline that highlights the most critical regions of an image used in a deep learning model's decision-making process, increasing interpretability and trust in the results. It is applied in many computer vision (CV) tasks such as classification and explanation. This study explores the principles of Explainable Deep Learning and its relevance to medical imaging, discusses various explainability techniques and their limitations, and examines medical imaging applications of Grad-CAM. The findings highlight the potential of Explainable Deep Learning and Grad-CAM in improving the accuracy and interpretability of deep learning models in medical imaging. The code is available in (will be available).
\end{abstract}
\section{Introduction}
\begin{figure}[htbp]
\centering
\includegraphics[width=\textwidth]{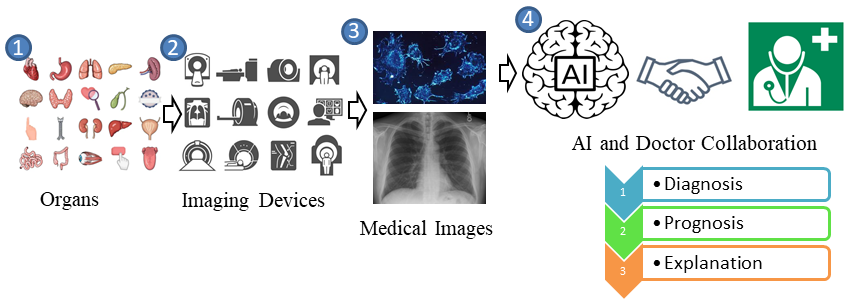}
\caption{A typical setup for explainable medical image analysis.}
\label{fig:method}
\end{figure}

Medical imaging, such as X-ray, CT, MRI, and ultrasound, plays a crucial role in the diagnosis and treatment of various diseases. With the increasing availability of medical imaging data, there is a growing interest in using machine learning techniques to aid in image analysis and interpretation. Fig. \ref{fig:method} shows a typical setup for explainable medical image analysis.

Deep learning has shown remarkable success in medical image analysis tasks, but one challenge with deep learning models is their lack of interpretability, which is a major concern in the medical domain. Explainable Deep Learning is an emerging area of research that seeks to address this issue by providing interpretable models without sacrificing accuracy. Explainable Deep Learning techniques aim to explain how a model arrived at its decision by providing saliency maps or highlighting the most relevant features in the input image that contributed to the model’s output. One popular Explainable Deep Learning technique is Gradient-weighted Class Activation Mapping (Grad-CAM), which visualizes the regions of an image that contribute most to the model’s decision-making process. Grad-CAM has been successfully used in various medical imaging applications, such as detecting lung cancer \cite{Wang2021LungCancer}, breast cancer \cite{Wang2021BreastCancer}, and brain tumors, to enhance the interpretability of deep learning models and improve trust in their results \cite{Selvaraju2017GradCAM}. This paper explores the use of Grad-CAM as an Explainable Deep Learning technique in medical imaging applications, discussing its underlying principles, how it works, and recent studies that have used Grad-CAM to improve the interpretability of deep learning models in medical imaging. Deep learning models lack transparency in their decision-making process, which is crucial for ensuring patient safety in the medical domain. The study proposes using Gradient-weighted Class Activation Mapping (Grad-CAM) as an Explainable Deep Learning technique to enhance interpretability. Grad-CAM allows clinicians to visualize the regions of an image that contribute most to the model's decision-making process, enabling better understanding and trust in the model's results.

The specific objectives of this paper are:

\begin{itemize}
    \item To review the literature on medical imaging and deep learning, explainable deep learning, and Grad-CAM in medical imaging such as the articles reported in \cite{shen2020deep,zeiler2014visualizing,samek2019towards}.
    \item To develop and compare state-of-the-art deep learning model for medical image interpretation \cite{wang2021deep}.
    \item To critically analyze Grad-CAM as an Explainable Deep Learning technique, evaluate its performance that can lead towards enhancing the interpretability.
    \item To discuss the implications and potential future applications of the Grad-CAM in different clinical settings such as \cite{rajpurkar2018deep,beam2016clinical}.
\end{itemize}

This paper contributes to the ongoing effort to address the lack of interpretability in deep learning models, particularly in the medical domain. By improving the interpretability of deep learning models, clinicians can better understand and trust the decision-making process of these models, which can improve patient safety and lead to more efficient and effective diagnosis and treatment of various diseases. The proposed pipeline can also potentially be applied in other fields where deep learning is used for image analysis.

\label{intro}

\section{Literature Review}

\subsection{Medical Imaging and Deep Learning}

A branch of machine learning called deep learning (DL) is able to learn to use its own computational strategy. Similar to how people make judgments, a deep learning model is employed to consistently organize data into a homogeneous framework \cite{esteva2019guide}. This is accomplished using deep learning, which layers a number of algorithms into a structure known as an artificial neural system (ANN). Medical imaging produces vast amounts of data, making it challenging and time-consuming for human radiologists to analyze effectively. Deep learning techniques can recognize patterns in medical images, identify anomalies, and categorize structures such as tumors or lesions, leading to faster and more accurate diagnoses and improved patient treatment regimens. However, the lack of annotated medical images for deep learning algorithm training and the potential for overfitting are significant challenges. Deep learning algorithms have been successful in diagnosing Alzheimer's and Parkinson's, locating malignant tumors, and find anomalies in MRI scans. In one study, deep learning algorithms outperformed human radiologists in detecting breast cancer in mammograms, with an accuracy rate of 94.5\% \cite{Wang2016DeepLearning}.

\subsection{Explainable Deep Learning}

Deep learning models are based on neural networks that are composed of multiple layers of interconnected artificial neurons \cite{Goodfellow2016}. These models are trained on large amounts of data and can learn complex patterns and relationships. However, these models can be considered as black boxes since their decision-making process is not transparent, and it is challenging to understand how they arrive at their predictions or decisions \cite{Murdoch2019}.

Explainable deep learning techniques provide transparency and interpretability to complex deep learning models. In medical imaging, these techniques allow clinicians to understand how models make decisions, improving accuracy and efficiency. Explainable deep learning is promising in cancer detection and personalized medicine, identifying patterns and relationships that predict treatment effectiveness.

Explainable deep learning techniques provide transparency, trust, and fairness in deep learning models, increasing their effectiveness and reliability \cite{samek2017explainable}. Activation maps, LIME, SHAP, and decision trees are commonly used techniques. However, these techniques can increase complexity, training time, and cost, and may compromise performance for interpretability. Some models, such as those using unsupervised or reinforcement learning, may be challenging to interpret. It is essential to consider the specific requirements of the model before deciding to use explainable deep learning techniques to make informed decisions and achieve optimal results \cite{ribeiro2016should}.

\subsection{Gradient-weighted Class Activation Mapping (Grad-CAM)}

Class Activation Mapping (CAM) is a method for visualizing the regions in an image that are important for a given classification task. It was first introduced by Zhou et al. in 2016 \cite{zhou2016learning} and has since become a popular tool for interpreting the predictions of deep convolutional neural networks (CNNs). Since the introduction of CAM, there has been a large amount of subsequent work in this field. Researchers have proposed a number of modifications and extensions to the original method, including the use of Grad-CAM (Gradient-weighted Class Activation Mapping), which uses the gradient of the output of the network with respect to the activations of the feature maps to generate the heatmap \cite{Selvaraju2017GradCAM}.

\begin{figure}[htbp]
\centering
\includegraphics[width=0.9\textwidth]{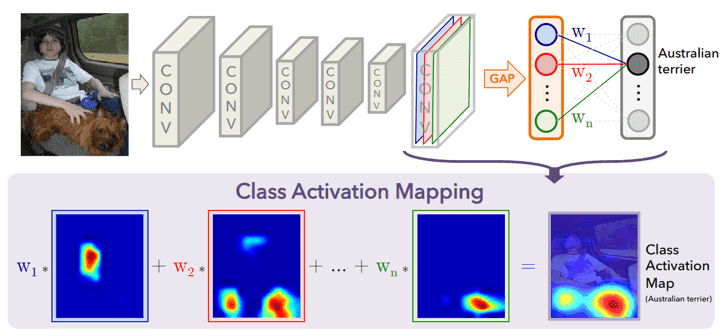}
\caption{Class Activation Mapping}
\label{fig:camelyon_low_resolution}
\end{figure}

Grad-CAM is a technique for visualizing where a CNN model is looking. It is class-specific and can produce a separate visualization for every class present in the image. It uses "alpha values" computed based on gradients to weight feature maps to create a heatmap. Grad-CAM has been used for cancer detection, and studies have shown that it improves accuracy and interpretability of results. However, limitations include lack of robustness to changes in input image and unclear explanation of the basis for prediction in complex images.

In recent years, there have been numerous efforts to improve the accuracy and interpretability of Grad-CAM for cancer detection\cite{wang2021}. One of the main focuses has been on improving the robustness of Grad-CAM to changes in the input image, as well as incorporating prior knowledge into the method to reduce the risk of misdiagnosis. One approach to improve the robustness of Grad-CAM is to use more sophisticated methods for computing the gradients used in the method\cite{qin2021}.

Incorporating prior knowledge into Grad-CAM is another approach to improving the interpretability of the results. For example, Li et al. proposed a method for incorporating prior knowledge into Grad-CAM that improved the accuracy of cancer detection in biopsy images. The authors showed that their proposed method reduced the risk of misdiagnosis and improved the interpretability of the results \cite{li2021incorporating}.

In addition to these efforts to improve Grad-CAM for cancer detection, there have also been efforts to extend the method to other medical imaging modalities, such as magnetic resonance imaging (MRI) and positron emission tomography (PET) scans. For example, Wang et al. proposed a method for using Grad-CAM for cancer detection in MRI scans, which showed promising results in a preliminary study \cite{wang2021grad}.

Overall, the recent advances in improving Grad-CAM for cancer detection demonstrate the potential of the method for improving the accuracy and interpretability of cancer detection in medical imaging. By addressing the limitations of the current methods and incorporating prior knowledge into the method, researchers are working towards making Grad-CAM a more reliable tool for cancer detection and diagnosis.

\section{Methodology}

We present the methodology used in our research to develop an accurate and interpretable model for the binary image classification task of identifying the presence of metastases from 96 x 96px digital histopathology images.

\bigskip

Lymph nodes are crucial in cancer diagnosis and staging, but the current diagnostic procedure for pathologists is tedious and time-consuming. However, ongoing advancements in technology and techniques offer hope for more accurate and efficient assessments in the future. Histopathological images of lymph nodes are essential for the diagnosis of various diseases, including cancer. The dataset used in this study comprises of 220,000 training images and 57,000 evaluation images, which is a carefully curated subset of the PCam dataset \cite{veeling2018} \cite{ehteshami2017diagnostic} derived from a comprehensive collection of H\&E stained whole slide images. The quality of the data is ensured through manual inspection by experienced technicians and consultation with pathologist. Specifically, there are 130,000 negative images and only 90,000 positive images, resulting in a ratio closer to 60/40. Overall, the quality, quantity, and recency of the data are sufficient, and the dataset meets our expectations. These advancements will benefit pathologists and improve patient outcomes by enabling earlier detection and more targeted treatment.

\bigskip

Identifying metastases in lymph nodes is a challenging task, even for trained pathologists. It can be especially difficult for untrained individuals due to the varied features that metastases exhibit. Irregular nuclear shapes, sizes, and staining shades are some of the indications of metastases. To train a model for classification, we must consider the optimal crop size for the images. We can avoid overfitting by increasing the amount of data, using augmentation techniques, implementing regularization, and simplifying model architectures. We can incorporate augmentation techniques like random rotation, crop, flip, lighting, and Gaussian blur into the image loader function to improve data quality. We must also examine samples with extremely low or high pixel intensities to ensure data quality. All bright images in the dataset have been labeled as negative, but we are unsure about the dark images. Hence, removing these dark images (outliers) from a large dataset is unlikely to affect prediction performance.

When building a machine learning (ML) production pipeline, it's best to start with a simple model to identify any issues, such as poor data quality, before tuning further. Maintaining equal ratios of negative and positive labels in training and validation sets is important to prevent under representation of rare classes. Selecting the ideal model architecture requires considering various factors, including the risk of overfitting with a deeper architecture. A pre-trained convnet model, densenet169, with transfer learning, is an effective approach for this problem. We use the one cycle policy for disciplined hyperparameter selection, which can save time spent training suboptimal hyperparameters.

\begin{figure}[htbp]
\centering
\includegraphics[width=0.7\textwidth]{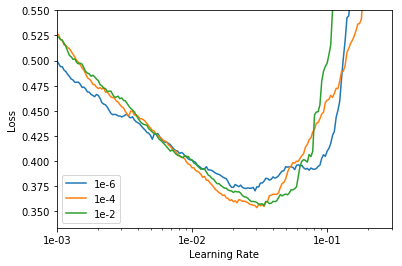}
\caption{Weight Decay Comparison}
\label{fig:weight_decay_comparison}
\end{figure}

First, we determine the optimal learning rate and weight decay values. The ideal learning rate is found just before the loss plateaus and divergence occurs, and it should be selected at a point where the loss is still decreasing. For weight decay, we conduct a small grid search using values of 1e-2, 1e-4, and 1e-6 (Figure \ref{fig:weight_decay_comparison}). Our goal is to find the highest weight decay that results in a low loss and allows us to use the maximum learning rate before it spikes. Based on experimentation, we found that 1e-4 is the largest weight decay that enables us to train with the maximum learning rate. We select a learning rate of around 2e-2, which is close to the bottom but still descending. A two-step training approach, beginning with the heads and freezing remaining parts, is used to optimize model performance. Unfreezing all trainable parameters allows fine-tuning the pre-trained bottom layers to detect common shapes and patterns, avoiding overfitting, and achieving better accuracy and robustness.

Now we describe the implementation of Grad-CAM (Gradient-weighted Class Activation Mapping) for visualizing and localizing the image regions that contribute the most to the model's predictions. The Grad-CAM method is implemented using the fastai library's hooks feature, which allows us to attach functions to a model's forward and backward passes. First, we define the \verb|hooked_backward| function that hooks into the convolutional part of the model and returns the activations and gradients. We use these activations and gradients to compute the Grad-CAM heatmap. To obtain the heatmap, we create a utility function \verb|getHeatmap| that takes a validation set image and returns the activation map. This function hooks into the model's forward and backward passes using the \verb|hooked_backward| function and computes the Grad-CAM map. The \verb|getHeatmap| function uses OpenCV to convert the batch tensor image to a grayscale image, which is then overlaid with the Grad-CAM heatmap using matplotlib.

Finally, we plot the Grad-CAM heatmap for a few selected validation images as shown in figure . We use the \verb|getHeatmap| function to obtain the heatmap and overlay it on top of the grayscale image. The resulting plots show the original image, the grayscale image, and the Grad-CAM heatmap overlaid on top. Hence, the Grad-CAM implementation provides a useful tool for understanding the decision-making process of a neural network and visualizing the image regions that contribute the most to its predictions as we will see in the next section.

\section{Results \& Discussion}

\subsection{Model Performance}

In this section, we present the results of evaluating the performance of our proposed model. 

\begin{figure}[htbp]
\centering
\includegraphics[width=0.7\textwidth]{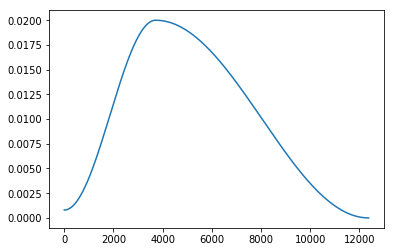}
\caption{Learning Rate of Initial Cycle}
\label{fig:initial_learning_cycle}
\end{figure}

Figure \ref{fig:initial_learning_cycle} displays the learning rate during the initial cycle of training. It starts at a low value and gradually increases to reach the maximum learning rate in the middle of the cycle, and then slows down towards the end. This approach uses a low warm-up learning rate initially, gradually increasing it to a higher value. The higher learning rate has a regularizing effect, preventing the model from settling for sharp and narrow local minima, and instead encouraging wider and more stable ones. As we approach the middle of the cycle, the learning rate is lowered to search for stable areas where minima may exist. This helps prevent overfitting and ensures the model can generalize well to new data.

\begin{figure}[htbp]
\centering
\includegraphics[width=0.7\textwidth]{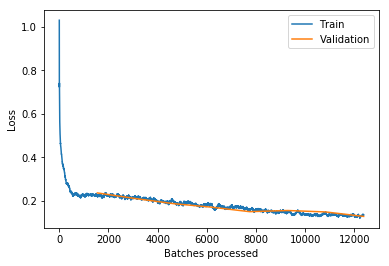}
\caption{Losses of Initial Cycle}
\label{fig:initial_losses}
\end{figure}

In figure \ref{fig:initial_losses}, we show the losses during the initial cycle of training. Upon analysis, we observe a slight increase in losses following the initial drop, which can be attributed to the escalating learning rate in the first half of the cycle. It is noteworthy that there is a temporary surge in losses when the maximum learning rate pushes the model out of local minima. However, this strategic move is expected to yield positive results as the learning rates are subsequently reduced.

\begin{figure}[htbp]
\centering
\includegraphics[width=0.7\textwidth]{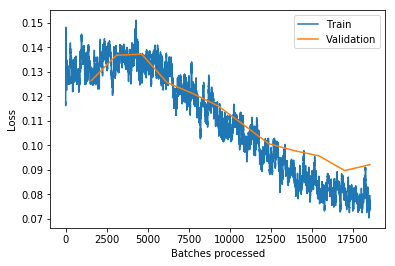}
\caption{Loss vs Batches Processed}
\label{fig:losses_vs_batches_processed}
\end{figure}

In Figure \ref{fig:losses_vs_batches_processed}, we show the losses from training and validation for batches processed. Upon closer examination, we observe that the validation performance has diverged from the training performance towards the end of the cycle. This indicates that our model has begun to overfit during the small learning rates. Further training would result in the model solely memorizing features from the training set, leading to an increase in the validation set performance. Therefore, it is a good stopping point for the training process.

\begin{figure}[htbp]
\centering
\includegraphics[width=1\textwidth]{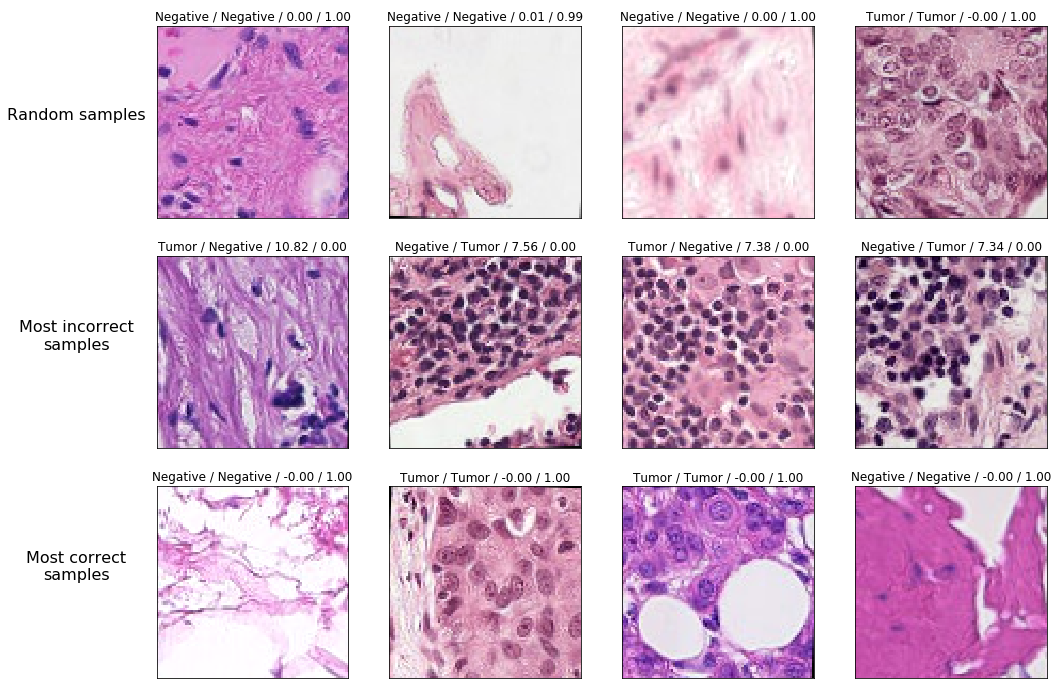}
\caption{Model Predictions (Predicted / Actual / Loss / Probability)}
\label{fig:model_results}
\end{figure}

Based on the validation set metrics, our model achieves an accuracy of 0.96735196\%. In Figure \ref{fig:model_results}, we evaluate our model's performance on different types of image examples. Firstly, we randomly select samples to observe the overall accuracy of our model. Next, we focus on the most incorrectly labeled images with high confidence to identify patterns and areas for improvement. Finally, we explore the most accurately labeled images to gain insights into our model's strengths. This visualization provides valuable insights into challenging images for the model, and may uncover issues with the dataset. By analyzing these examples, we can improve the model's performance and gain a better understanding of the data.

\subsection{Explainability of the Model with Grad-CAM}

\begin{figure}[htbp]
\centering
\includegraphics[width=1\textwidth]{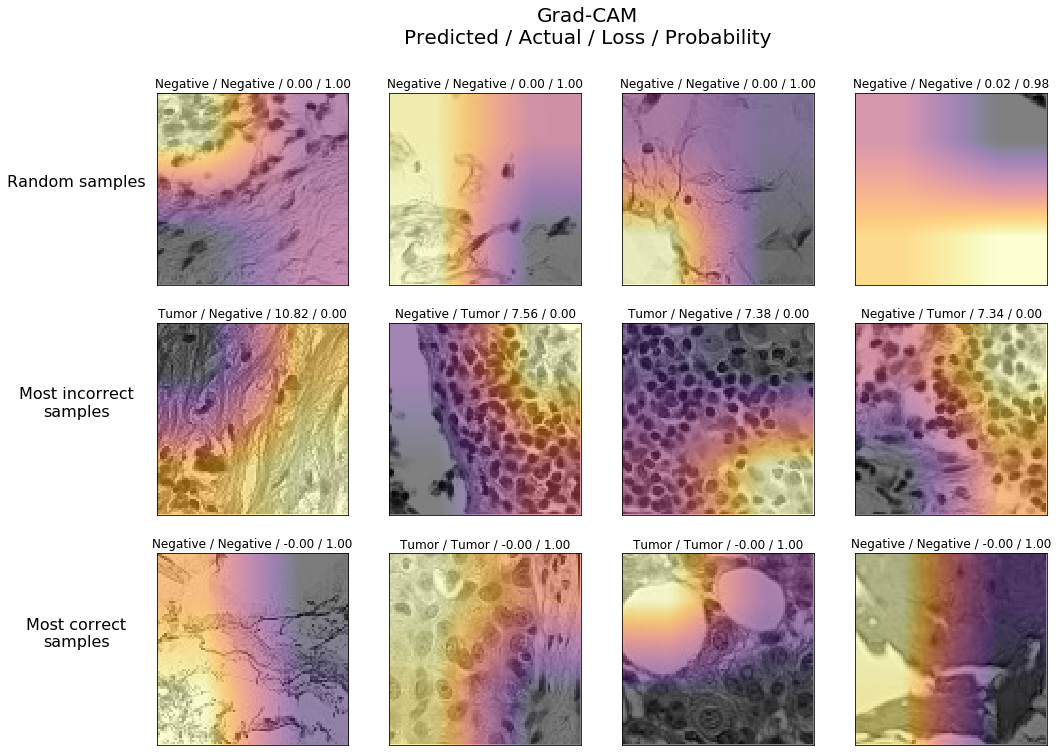}
\caption{Grad-CAM Results (Predicted / Actual / Loss / Probability)}
\label{fig:grad_cam_results}
\end{figure}

In this section, we explain the explainability of our model using Gradient-weighted Class Activation Mapping (Grad-CAM). Grad-CAM is a visualization technique that highlights important regions in the input image that contribute to the output class score. By applying Grad-CAM to our model, we gain insights into its decision-making process and identify important features for classification. To generate Grad-CAM visualizations, we select an image from the validation set, obtain the predicted class score, and compute gradients of the score with respect to feature maps. These gradients serve as importance weights for the feature maps, and we combine them to produce a heat map that highlights relevant regions in the input image. Figure \ref{fig:grad_cam_results} shows the Grad-CAM visualizations for different types of image examples. For instance, if the label is \emph{tumor}, Grad-CAM will reveal all the locations where the model believes the tumor patterns exist.

\section{Conclusion}
In conclusion, this paper has explored the potential of using Grad-CAM as a technique for building explainable deep learning models in medical imaging, with a focus on cancer detection. While the findings are promising, there is still much work to be done in this field. Future research could focus on accuracy of Grad-CAM visualisations, extending it to other imaging modalities, and developing user interfaces that provide clinicians with intuitive and interpretable visualizations of the model’s decision-making process. Overall, this paper highlights the importance of Explainable Deep Learning in medical imaging and the potential of Grad-CAM as a tool for enhancing interpretability and trust in deep learning models.

%
%
%

 \bibliographystyle{unsrt}
 \bibliography{library}
%

\end{document}